\documentclass[lnl,prl,a4paper,twocolumn]{revtex_lnl} 
 \usepackage{epsfig}
 \usepackage{graphicx}
\usepackage[none]{hyphenat} 
\usepackage{pslatex}
\usepackage{mathtools}
\usepackage{bm}
\usepackage{amsmath,amssymb}

\begin{document}


\onecolumngrid
\begin{center}
\begin{large}
{\bf Modeling Nuclear Reactions for PET/MRI MultiModal Imaging: the innovative use of $^{52g}$Mn}
\end{large}
\vspace{.4cm}

A. Colombi$^{1,2}$, F. Barbaro$^{3}$, L. Canton$^{3}$, M.P. Carante$^2$, A. Fontana$^2$
\vspace{.4cm}

\textit{\small
$^1$ Dipartimento di Fisica dell'Universit\`a di Pavia, Pavia, Italy. $^2$ INFN, Sezione di Pavia, Pavia, Italy.\\
$^3$ INFN, Sezione di Padova, Padova, Italy. \\
}
\end{center}
\vspace{0.65cm}

\thispagestyle{empty} 
\pagestyle{empty} 
\twocolumngrid

\section{Introduction}

MultiModal Imaging (MMI) is an emerging medical technique, aiming to increase the diagnostic image information by simultaneously using different physical processes:
a possible application is represented by a combined PET/MRI scan, by the use of a $\beta^+$ emitting radionuclide \space showing \space paramagnetic \space properties. An emerging 
radioisotope characterized by both these features is $^{52g}$Mn, which decays in $^{52}$Cr by emitting a 242 keV positron, with an half-life of about 5.6 days.

The development of a technique to produce $^{52g}$Mn at cyclotrons is the main goal of the METRICS (Multimodal pET/mRi Imaging with Cyclotron-produced $^{52/51}$ Mn
and stable paramagnetic Mn iSotopes) project, in the framework of the LARAMED project at INFN-LNL \cite{ref4}. One crucial step for this study is the identification of
the best nuclear reactions and irradiation parameters in order to produce $^{52g}$Mn with high purity. The theoretical approach developed for this purpose by the
Pavia and Padova groups is illustrated in this report.

\section{Routes to $^{\textbf{52g}}$M$\textbf{n}$}

The standard route for the production of $^{52}$Mn is represented \space by \space the \space reaction: $^{52}$Cr(p,n)$^{52g}$Mn \space \cite{ref1}. The adopted methods of the present work
and the obtained results are here shown for this reaction, though also many other channels have been investigated.

\begin{figure}[h]
\begin{center}
\includegraphics[width=\columnwidth]{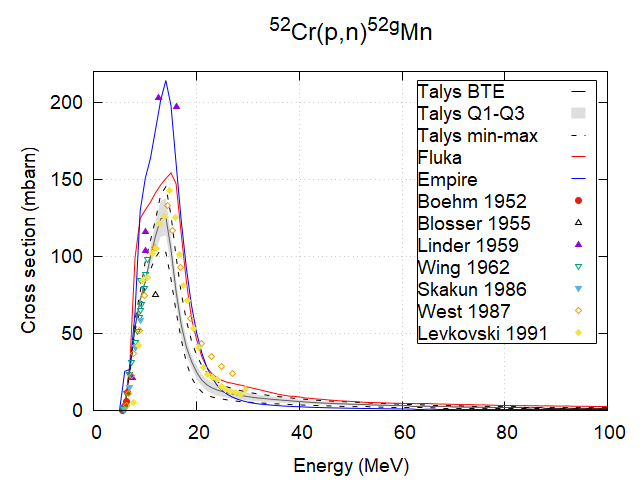}
\end{center}
\caption{Theoretical estimations and experimental data for the cross section of the $^{52}$Cr(p,n)$^{52g}$Mn reaction.}
\label{label_fig1}
\end{figure}

Theoretical calculations of cross sections have been performed by using three nuclear reaction codes: TALYS, FLUKA and EMPIRE. For TALYS, in particular, 4 
pre-equilibrium models and 6 level density models are implemented, leading to a total of 24 different model combinations. In \space Fig. \ref{label_fig1}, \space the \space calculated
\space cross \space sections of the aforementioned reaction for the different codes are shown: the grey band represents the region between the first and the third quartile
(Q1-Q3) of the 24 cross sections calculated with the different TALYS models, whereas the two dashed lines represent the minimum and maximum calculated values, and
the solid line is the "Best Theoretical Evaluation" (BTE) defined as the center of the interquartile range \space \cite{ref2}.
Along \space with \space the \space theoretical \space calculations, \space also several experimental data taken from the literature are shown. Despite some differences between the models and the
experimental data, a maximum of the cross section is evident between 10 and 20 MeV.

\begin{figure}[h]
\begin{center}
\includegraphics[width=\columnwidth]{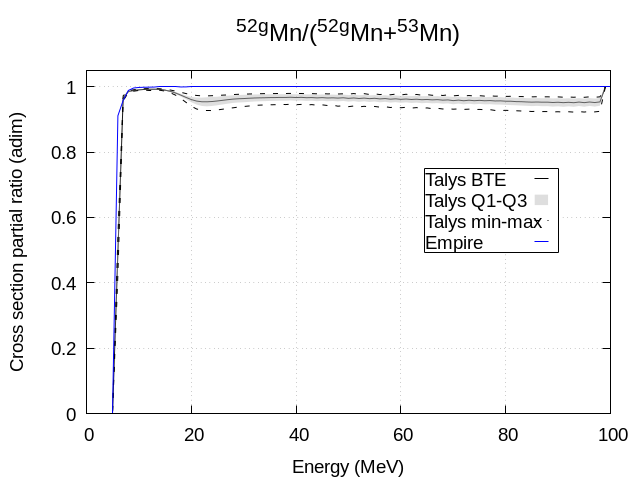}
\end{center}
\caption{Cross-section ratio between $^{52g}$Mn and its sum with $^{53}$Mn, for the $^{52}$Cr(p,n)$^{52g}$Mn reaction.}
\label{label_fig2}
\end{figure}

Since the majority of the produced manganese isotopes (namely $^{48}$Mn, $^{49}$Mn, $^{50g/m}$Mn, $^{51}$Mn and $^{52m}$Mn) are characterized by a half-life
shorter than one hour, their contamination, both in terms of isotopic and radionuclidic purity, \space would \space be \space negligible \space after \space few \space hours. It \space is \space thus interesting to
evaluate the amount of $^{53}$Mn, whose half-life is of about 3.6$\times 10^{6}$ years, that can be produced through this reaction, compared to that of $^{52g}$Mn.
In order to do that, Fig. \ref{label_fig2} shows the ratio
\begin{equation}
	r=\frac{\sigma_{^{52g}Mn}}{\sigma_{^{52g}Mn}+\sigma_{^{53}Mn}},
\end{equation}
where $\sigma$ is the cross section. The production of $^{52g}$Mn appears almost pure, with respect to its main contaminant, in the energy region between 10 and 20
MeV, where the maximum of its cross section occurs.

%
%
%
Starting from the obtained cross sections, a simulation of an irradiation experiment in the chosen energy window was performed. The production rates of all the
produced manganese isotopes were calculated as
\begin{equation}
	R = \frac{I_0}{z_{proj}|e|}\frac{N_a}{A}\int_{E_{out}}^{E_{in}}\sigma(E)\left(\frac{dE}{\rho_tdx}\right)^{-1}dE
\end{equation}
where $I_0$ is the charge beam current, $z_{proj}$ the atomic number of the incident particle, $e$ the electron charge, $N_a$ the Avogadro number,
$A$ the target atomic mass, $E_{in}$ and $E_{out}$ the energy of the projectile impinging on the target and after exiting from the target respectively, 
$\sigma(E)$ the production cross section of the considered nuclide, $\rho_t$ the target density and $dE/dx$ the stopping power of the projectile in the target. 
In this case the chosen parameters of the irradiation are a beam current of 1 $\mu A$, an incident energy of 17 MeV, a target thickness of 200 $\mu m$ 
(corresponding to $E_{out}=15$ MeV) and an irradiation time of 1 h. The temporal evolution of the number of nuclei N of the various isotopes was then computed, 
by means of standard Bateman equations.

\begin{figure}[h]
\begin{center}
\includegraphics[width=\columnwidth]{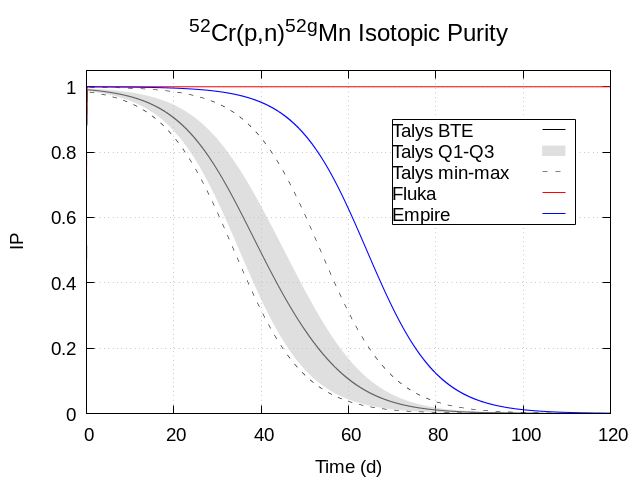}
\includegraphics[width=\columnwidth]{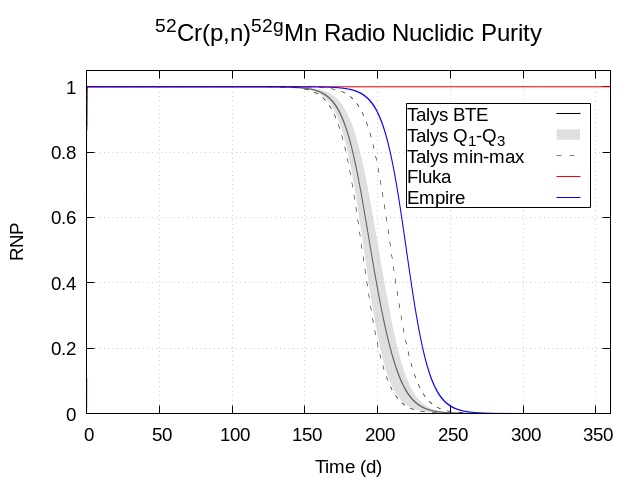}
\end{center}
\caption{Isotopic Purity (top panel) and Radionuclidic Purity (bottom panel) of $^{52g}$Mn for the $^{52}$Cr(p,n)$^{52g}$Mn reaction.}
\label{label_fig4}
\end{figure}

The Isotopic Purity (IP) and the RadioNuclidic Purity (RNP), are defined as
\begin{align}
\begin{split}
	IP(t) = \frac{N_{^{52g}Mn}}{\sum_{i}{N_{^{i}Mn}}}, \qquad
	RNP(t) = \frac{A_{^{52g}Mn}}{\sum_{i}{A_{^{i}Mn}}}
\end{split}
\end{align}
where N represents the number of nuclei and A the activity of the nuclide. These quantities are calculated as a function of time and shown in Fig. 
\ref{label_fig4}. The grey band of TALYS has a meaning analogous to
that described for Fig. \ref{label_fig1}, whereas for FLUKA the ratios are always equal to 1, because this code cannot calculate (p,$\gamma$) reactions, and the 
only long-life contaminant produced in the considered energy window is $^{53}$Mn, through the reaction: $^{52}$Cr(p,$\gamma$)$^{53}$Mn. All the codes
predict a IP greater than 0.8 for at least 20 days after the irradiation, and a RNP close to 1 for at least 150 days, that is a much longer time with respect to
the half life of $^{52g}$Mn. This means that there is not a significant contribution to the total activity from the other produced Mn isotopes, which would be
otherwise a problem for the patient health. For this reason, the (p,n) reaction is promising for the production of $^{52g}$Mn with high purity. 
Also the activity per unit of current at the End-of-Bombardment was calculated according to the different models, and its values in MBq/$\mu$Ah are: 
6.64 $\pm$ 1.7 for TALYS \cite{ref2}, 20.2 for EMPIRE and 23.2 for FLUKA.

%
%
%
%

The procedure described for this reaction has a general validity and was applied to many other reactions, in order to identify other possible competitive channels
for the production of high purity $^{52g}$Mn. The following reactions were in particular analyzed: $^{53}$Cr(p,2n)$^{52g}$Mn,  $^{54}$Cr(p,3n)$^{52g}$Mn,
$^{52}$Cr(d,2n)$^{52g}$Mn, $^{54}$Fe(p,$^3$He)$^{52g}$Mn, $^{54}$Fe(d,$\alpha$)$^{52g}$Mn, $^{56}$Fe(p,$\alpha$n)$^{52g}$Mn, $^{56}$Fe(d,$\alpha$2n)$^{52g}$Mn,
$^{Nat}$V($\alpha$,3n)$^{52g}$Mn. Among these, the (d,2n) and ($\alpha$,3n) reactions proved to be the most promising channels \cite{ref3}.

\section{Conclusions}

The goal of this work is to identify the best nuclear reactions and irradiation parameters for the production of $^{52g}$Mn with high purity, in view
of its clinical application in the field of MMI. Although further experimental data are needed to confirm the theoretical results, the adopted approach and the
developed tool are versatile and can be applied to the optimization of the production of other radionuclides, also for different practical applications.


\end{document}